\documentclass[twocolumn,showpacs,preprintnumbers,amsmath,amssymb,pre]{revtex4}

\usepackage{graphicx}
\usepackage{dcolumn}
\usepackage{bm}
\usepackage[]{epsfig}

\newcommand{\Nb}{N_{\mathrm {blob}}}
\newcommand{\Ne}{N_{\mathrm {ev}}}

\newcommand{\Vb} {V_{\mathrm b}}

\begin{document}
\title{Investigation of $q$-dependent dynamical heterogeneity in a
    colloidal gel by x-ray photon correlation spectroscopy}

\author{V. Trappe$^1$, E. Pitard$^2$, L. Ramos$^2$, A. Robert$^{3,4}$, H. Bissig$^1$, L.
Cipelletti$^{2,*}\email{lucacip@lcvn.univ-montp2.fr}$}
\affiliation{ $^1$Departement de Physique, Universit\'{e} de
Fribourg, Chemin du
Mus\'{e}e 3, 1700 Fribourg, Suisse\\
$^2$Laboratoire des Collo\"{\i}des, Verres et Nanomat\'{e}riaux (UMR
CNRS-UM2 5587), CC26, Universit\'{e}
Montpellier 2, 34095 Montpellier Cedex 5, France \\
$^3$European Synchrotron Radiation Facility - 6 rue Jules Horowitz
BP 220, F-38043 Grenoble Cedex 9, France\\
$^4$Present address: Stanford Linear Accelerator Center 2575 Sand
Hill Road, Menlo Park, CA 94025, USA }

\email{lucacip@lcvn.univ-montp2.fr}
\date{\today}

\begin{abstract}
We use time-resolved X-Photon Correlation Spectroscopy to
investigate the slow dynamics of colloidal gels made of moderately
attractive carbon black particles. We show that the slow dynamics is
temporally heterogeneous and quantify its fluctuations by measuring
the variance $\chi$ of the instantaneous intensity correlation
function. The amplitude of dynamical fluctuations has a
non-monotonic dependence on scattering vector $q$, in stark contrast
with recent experiments on strongly attractive colloidal gels [Duri
and Cipelletti, \textit{Europhys. Lett.} \textbf{76}, 972 (2006)].
We propose a simple scaling argument for the $q$-dependence of
fluctuations in glassy systems that rationalizes these findings.
\end{abstract}

\pacs{64.70.Pf, 82.70.Gg, 82.70.Dd}
\maketitle

\section{Introduction}
Understanding the dramatic slowing down of the dynamics in systems
undergoing a glass transition is one of the key problems in
condensed matter and statistical physics~\cite{Donth2001}. In recent
years, research efforts have focused on the role of dynamical
heterogeneity: as the glass transition is approached, the dynamics
becomes increasingly correlated in space, since rearrangements are
possible only through the cooperative motion of ``clusters'' of
particles ~\cite{EdigerReview,RichertReview,GlotzerReview}. This
cooperativity leads to strong temporal fluctuations of the dynamics.
Indeed, because of dynamical correlations, the number of
statistically independent objects in the system becomes smaller than
the number of particles, leading to enhanced fluctuations.

Numerical simulations have tested these features on a wide variety
of systems~\cite{GlotzerReview}. Experimental work, by contrast, is
much more scarce, because probing the dynamics with the spatial and
temporal resolution needed to highlight their heterogeneous nature
is an arduous task, especially for molecular glass
formers~\cite{EdigerReview,RichertReview}. The slow dynamics of
colloidal systems, foams and granular materials~\cite{LucaJPCM2005}
share intriguing similarities with those of glass formers, including
dynamical heterogeneities. These are experimentally more accessible
than in molecular systems, since the relevant time and length scales
are larger. Various techniques have been used to characterize them,
from direct space measurements
~\cite{WeeksScience2000,DauchotPRL2005_2,KeysNaturePhysics2007
} to novel scattering methods that probe the temporal fluctuations
of the intensity correlation
function~\cite{LucaJPCM2003,MayerPRL2004,SarciaPRE2005,
DuriEPL2006}.

Valuable information on the physical origin of the average dynamics
is generally obtained by studying its length scale dependence, e.g.
the dependence of the intensity correlation function on the
magnitude of the scattering vector $q$ in scattering experiments.
Similarly, one expects to gain a better understanding of dynamical
heterogeneities by analyzing their behavior at different $q$'s.
Unfortunately, experimental and numerical or theoretical
determinations of dynamical fluctuations as a function of $q$ are
still very
scarce~\cite{DauchotPRL2005_2,DuriEPL2006,ChandlerPRE2006,BBBKMR,LacevicJChemPhys2003,CharbonneauCondmat2007,AbetePRL2007},
leaving this issue an open question.

In this paper, we investigate dynamical fluctuations in colloidal
gels made of moderately attractive carbon black (CB) particles, to
which a dispersant is added to control the strength of the
(attractive) interparticle interactions. We apply, to our knowledge
for the first time, time-resolved scattering methods to X-Photon
Correlation Spectroscopy (XPCS), thereby demonstrating that the
dynamics of the CB gels are temporally heterogeneous. Dynamical
fluctuations are quantified by means of a $q$-dependent dynamical
susceptibility, $\chi$, similar to the dynamical susceptibility
$\chi_4$ studied in simulations~\cite{LacevicJChemPhys2003}.
Surprisingly, $\chi$ is found to initially increase with $q$, but
eventually to decrease at large scattering vectors. This
non-monotonic behavior is in contrast with recent low-$q$
measurements on diluted, strongly attractive colloidal
gels~\cite{DuriEPL2006}, where dynamical fluctuations increased
linearly with $q$ over one decade in scattering vector. We propose a
simple scaling argument for the $q$-dependence of dynamical
fluctuations in glassy systems, which reconciles these contrasting
findings and rationalizes previously published data for granular
media and glass
formers~\cite{DauchotPRL2005_2,ChandlerPRE2006,BBBKMR,CharbonneauCondmat2007,LacevicJChemPhys2003}.

The paper is organized as follows: in Sec.~\ref{Sec:MM} we present
and characterize our experimental system and introduce shortly the
time-resolved XPCS technique. Section~\ref{Sec:R} reports both the
average dynamics of the CB gels and its temporally fluctuations. Our
results are discussed in Sec.~\ref{Sec:D}, where a simple scaling
argument for the length scale dependence of the dynamical
susceptibility is introduced.

\section{Materials and methods}
\label{Sec:MM}
\subsection{Sample preparation and characterization}

\textit{Particle size and morphology.} Our gels are made of CB
particles of average radius $\overline{R} = 180$ nm, suspended in
mineral oil at an effective volume fraction $\varphi \approx 6\%$,
as determined by viscosity measurements (see below). The particles
have a conveniently high scattering cross section for X-rays. They
are fractal aggregates made of permanently fused primary particles
(fractal dimension $d_\mathrm{f}
 = 2.2 \pm 0.1$)~\cite{TrappePRL2000}. The diameter of the primary particles ranges from 20 to
40 nm, as determined by electron microscopy. However, the smallest
units in our samples are effectively the CB particles themselves and
not the primary particles, since the CB particles can not be broken,
neither by thermal fluctuations, nor by adding a dispersant or by
applying a large shear, e.g. during rheology tests or sonication.
The size and polydispersity of the CB particles were determined by
applying the cumulant analysis described in
Ref.~\cite{PuseyJChemPhys1984} to intensity correlation functions
measured by dynamic light scattering (DLS) at $q =
6.01~\mu\mathrm{m}^{-1}$ (scattering angle $\theta = 20$ deg). A
very diluted ($\varphi \approx 2 \times 10^{-6}$) and fully
dispersed sample of CB-particles in mineral oil was prepared for the
DLS measurements. For spherical, monodisperse particles, one expects
$g_2-1$ to relax exponentially; deviations due to shape and/or size
polydispersity may be quantified by the ratio $\kappa_2/
\kappa_1^2$, where $\kappa_i$ is the $i-$th coefficient of a
cumulant expansion: $\ln\{[g_2(\tau)-1]^{0.5}\} = \kappa_0 -
\kappa_1\tau + \kappa_2\tau^2/2 + ...$~\cite{PuseyJChemPhys1984}.

\begin{figure}
   \epsfig{file=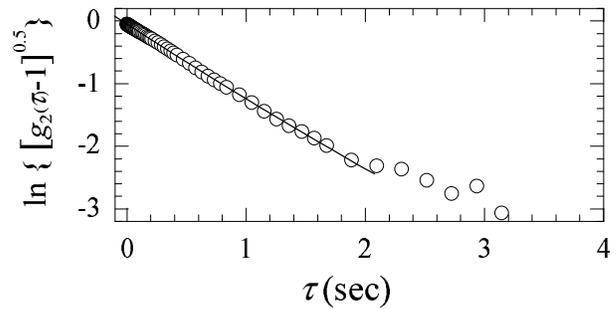,width=8cm}
 \caption{Second-order cumulant fit (line) of the intensity correlation function (open circles)
 measured by dynamic light scattering at
 $q =6.01~\mu\mathrm{m}^{-1}$ for a diluted suspension of CB particles.}
 \label{Fig:1}
\end{figure}

Figure~\ref{Fig:1} shows $\ln\{[g_2(\tau)-1]^{0.5}\}$ $vs$ $\tau$
for our DLS measurements: the data are very close to a straight
line, indicating a nearly exponential decay and thus suggesting that
the sample polydispersity must be moderate. By generalizing the
arguments of Ref.~\cite{PuseyJChemPhys1984} to particles with a
fractal morphology, one finds that in the low-$q$ limit ($qR
\lesssim 1$) the first two cumulants are related to the average
translational diffusion coefficient, $\overline{D}$, and its
relative variance, $\sigma^2_{D} =
(\overline{D^2}/\overline{D}^{\,2}-1)$, by the following
expressions:

\begin{eqnarray}
   & \kappa_1 = \overline{D}q^2
\\
 & \kappa_2 = \sigma^2_{D}(\overline{D}q^2)^2/2\, ,
     \label{equ:Cndef}
\end{eqnarray}
with $\overline{D} = k_{\mathrm B}T/(6 \pi \eta R_{\mathrm {app}})$,
$R_{\mathrm {app}} =
\overline{R^{2d_\mathrm{f}}}/\overline{R^{2d_\mathrm{f}-1}}$, $T$
the temperature and $k_{\mathrm B}$ the Boltzman's constant. Here,
$\overline{R^n} = \int \mathrm{d}R P(R)R^n$ is the $n$-th moment of
the normalized number distribution of particle radii, $P(R)$. For
the data shown in Fig. 1, we find $\sigma^2_{D} =
2\kappa_2/\kappa_1^2 = 0.047$, confirming that the polydispersity is
moderate. The average particle radius and its relative variance,
$\sigma^2_{R} = (\overline{R^2}/\overline{R}^{\,2}-1)$, may be
obtained from $\sigma^2_{D}$ and $R_{\mathrm {app}}$ provided that
$P(R)$ is known. Since for moderate polydispersity the exact shape
of $P(R)$ has little influence on the final result, we choose a
generalized exponential (or Schulz) distribution, for which
calculations can be performed
analytically~\cite{PuseyJChemPhys1984}. By taking for simplicity
$d_\mathrm{f}  = 2$ (very close to the value 2.2 of our particles),
one finds
\begin{eqnarray}
   R_{\mathrm {app}} = \overline{R^4}/\overline{R^3} =
   \overline{R}(1+3\sigma^2_R)
\\
\sigma^2_{D} = \frac{\overline{R^2}
\,\,\overline{R^4}}{\overline{R^3}^{\,2}}-1 =
\frac{1+3\sigma^2_R}{1+2\sigma^2_R}-1 \, ,
     \label{equ:Cndef}
\end{eqnarray}
yielding for our CB particles $\overline{R} = 180$ nm and $\sigma_R
= 0.23$.\newline

\textit{Determination of the volume fraction} The particle
concentration of the two CB gels studied here is 2\% w/w. The
effective volume fraction corresponding to this weight fraction was
determined by measuring the viscosity of a suspension where the CB
particles were fully dispersed (1.6\% w/w dispersant). The effective
hydrodynamic volume fraction, $\varphi$, is estimated
using~\cite{Russel} $\eta_{\infty}/\mu = (1-\varphi/0.71)^{-2}$,
where $\mu$ is the shear-independent viscosity of the mineral oil
and $\eta_{\infty}$ is the high shear rate viscosity of the
suspension. The value $\varphi \approx 6\%$ thus obtained was found
to be consistent with that obtained by measuring the low-shear
viscosity of the suspension, $\eta_0$, and using~\cite{Russel}
$\eta_{0}/\mu = (1-\varphi/0.63)^{-2}$.

\textit{Estimate of the depth of the particle-particle interaction
potential.} The attraction between CB particles is controlled by the
amount of added dispersant. The depth of the well of the
particle-particle interaction potential, $U$, may be estimated by
studying the mechanical response of the CB gels. As discussed in
ref.~\cite{PrasadFaraday2003}, the relaxation spectrum (as measured
by oscillatory rheology) of CB suspensions and its dependence on
volume fraction is essentially identical to that obtained for model
systems of nearly monodisperse spherical particles that interact via
a short-ranged potential induced by the depletion mechanism.
Therefore, we use the non-equilibrium state diagram of the depletion
systems with known depth of the interaction potential to evaluate
$U$ for our CB gels. The diagram is shown in Fig. 3 of
Ref.~\cite{PrasadFaraday2003}: the control parameters are $\varphi$
and $U$; a line $U = U_c(\varphi)$ separates the systems with
solid-like mechanical behavior (upper part of the diagram) to those
with a liquid-like behavior (lower part). For the CB gel with 0.2\%
dispersant, we find that the elastic modulus is very low ($G_p = 9
\times 10^{-4}$ Pa), which indicates that this system corresponds to
a state close to the fluid-solid boundary. Figure 3 of
Ref.~\cite{PrasadFaraday2003} indicates that for short ranged
potentials the boundary corresponds to $U = U_c \approx
12k_{\mathrm{B}}T$ at $\varphi = 6\%$. Thus, we assign $U \approx
12k_{\mathrm{B}}T$ to the CB gel at $\varphi = 6\%$ and with 0.2\%
dispersant.

For the CB-system with 0\% dispersant, we take advantage of the
dependence of the elastic modulus on dispersant concentration, which
was found to exhibit a critical-like behavior, $G_p =
G_0(U/U_c-1)^{\nu}$, with $\nu = 3.9$ and $G_0 \approx 1$
Pa~\cite{TrappePRL2000}. We measure $G_p = 4.5$ Pa for the CB gel
without dispersant, yielding $U \approx 2.5 U_c \approx
30~k_{\mathrm{B}}T$.

\textit{Sample preparation.} The CB suspensions are prepared from a
stock suspension of 4\% w/w carbon black (Cabot Vulcan XC72) in
light mineral oil (Aldrich). The stock solution is made by first
mixing the carbon black powder into the mineral oil using a standard
household mixer. Subsequently, it is thoroughly sonicated for ~1
hour using an ultrasound device (Hielscher UP200H) operating at a
power of 120 W and a frequency of 24 kHz. To avoid overheating, the
suspension is placed in an ice bath and sonication is run in cycles
of 1 s. The stock solution is then left to equilibrate for 2 days,
before it is used to prepare the final samples. Before diluting the
sample to a final concentration of 2\% w/w CB (corresponding to
$\varphi \approx 6 \%$), the stock suspension is again thoroughly
mixed; the final samples are then prepared by adding either pure
mineral oil or a solution of dispersant in mineral oil; they are
mixed and left to equilibrate for 2 days. Prior to the XPCS
experiments, the samples are again thoroughly mixed and subsequently
injected in a 1 mm (inner diameter) cylindrical capillary. The
capillary is sealed and placed in a water bath in which we introduce
our ultrasound device, thereby submitting the sample to a final
indirect sonication step (120 W and 24 kHz) for about half an hour.
We define the age of sample, $t_w$, as being the time elapsed since
this final sonication step.  Note that sonication breaks up any
aggregates that may have been formed between the CB particles,
without affecting the integrity of the CB particles themselves.

\subsection{Time-resolved XPCS measurements}

\textit{Average dynamics.} The dynamics of the CB gels are
investigated by means of XPCS performed at the ID10 Troika beamline
at the European Synchrotron Radiation Facility (ESRF), with X-rays
of wavelength $\lambda = 1.55$ {\AA}. The scattering volume has a
cylindrical shape, with diameter 12 $\mu$m and length 1 mm (along
the horizontal direction), defined by the beam size and the
capillary diameter, respectively. The scattered intensity is
recorded by a charge-coupled device (CCD) camera, covering about one
decade in scattering vector: $15~\mu \mathrm{m}^{-1} \lesssim q
\lesssim 150~\mu \mathrm{m}^{-1}$, corresponding to distances
comparable to or smaller than the CB particle size. XPCS
measurements start at age $t_w = 2$ min: we find that the dynamics
initially slows down, as observed for many glassy
systems~\cite{LucaJPCM2005}, but that after a few hours a stationary
state is attained. All measurements presented in this work start at
$t_w = 5$ hours, well in the stationary regime, and last typically
up to 3 hours. Mechanical and beam instabilities are a concern when
measuring very slow dynamics, especially in XPCS experiments. By
using a static scatterer (Vycor glass), we have checked that the
setup was stable up to $\tau \approx 3000$ sec, longer than the
relevant time scales in our experiments.

In order to measure the average dynamics and its fluctuations, we
use the time-resolved correlation (TRC) method~\cite{LucaJPCM2003}.
The instantaneous degree of correlation between X photons scattered
at time $t$ and $t+\tau$ is measured according to $c_I(q,t,\tau) =
\left< I_p(t)I_p(t+\tau) \right >_q / \left( \left <I_p(t) \right
>_q \left <I_p(t + \tau ) \right >_q \right) - 1$, where $I_p(t)$
is the scattered intensity at pixel $p$ and time $t$ and $\left <
\cdot \cdot \cdot \right>_q$ is an average over a ring of pixels
corresponding to approximately the same magnitude of $\mathbf{q}$
but different azimuthal orientations. The intensity correlation
function is $g_2(q,\tau)-1 = \overline{c_I}$, where
$\overline{\cdot\cdot\cdot}$ indicates a time average. $g_2$ is
related to the dynamic structure factor $f(q,\tau)$ by $g_2-1 =
\beta f^2$, where $\beta < 1$ is a positive instrumental constant.

\textit{Fluctuations of the dynamics.} To quantify dynamical
heterogeneity we calculate $\chi(\tau,q)^{\mathrm{(exp)}} =
\mathrm{var}(c_I(q,t,\tau))$, the temporal variance of the
instantaneous degree of correlation,
$c_I(q,t,\tau)$~\cite{notenormalization}, which we correct by
subtracting the contribution of measurement noise. The correction
procedure is described in detail in Ref.~\cite{DuriPRE2005}; here we
simply recall its main features. The experimentally measured degree
of correlation is affected by a statistical noise stemming from the
finite number of pixels, $n_{\mathrm{p}}$, over which this quantity
is averaged. Therefore, one has~\cite{DuriPRE2005}
\begin{equation}
\nonumber \chi(\tau,q) = \chi^{\mathrm{(exp)}}(\tau,q)
-\mathrm{var}[n(q,t,\tau)] = \chi^{\mathrm{(exp)}}(\tau,q) -
C/n_{\mathrm{p}} \, ,
\end{equation}
with $\chi^{\mathrm{(exp)}}(\tau,q) = \mathrm{var}[c_I(q,t,\tau)]$,
$\chi(\tau,q)$ the desired noise-free dynamical susceptibility,
$n(q,t,\tau)$ the statistical noise, and $C$ a positive constant.
The last equality stems from the central limit theorem. In order to
obtain $\chi(\tau,q)$, for each $q$ we calculate
$\chi^{\mathrm{(exp)}}(\tau,q)$ by processing the same set of images
with different choices of $n_{\mathrm{p}}$ (by using only 1 pixel
every 1, 2, 4, 8,... available pixels). By extrapolating the
$\chi^{\mathrm{(exp)}}(\tau,q)$ \textit{vs} $1/n_{\mathrm{p}}$ data
to $1/n_{\mathrm{p}} = 0$, we obtain the desired noise-free
$\chi(\tau,q)$. To estimate the uncertainty associated with this
procedure, we inspect the data for $\tau$ much larger than the
slowest relaxation of $g_2-1$, where one expects $\chi(\tau,q) =
0$~\cite{MayerPRL2004,DuriPRE2005}. In this limit, we indeed find
$\chi = 0$ (see Figs.~2b and~3b below) within an experimental
uncertainty, $\sigma_{\chi}(q)$, which is estimated by calculating
the standard deviation of $\chi(\tau,q)$ as a function of $\tau$, at
large $\tau$. The values of $\sigma_{\chi}(q)$ thus obtained are
taken as an estimate of the error bars on $\chi$ at all delays, and
will be used in particularly in Fig.~5. The average relative
uncertainty is 6\% for the CB gel with $U \approx
12~k\mathrm{_{B}}T$ and less than $1\%$ for the gel with $U \approx
30~k\mathrm{_{B}}T$.

\section{Results}
\label{Sec:R}

\begin{figure}
\centerline{\includegraphics{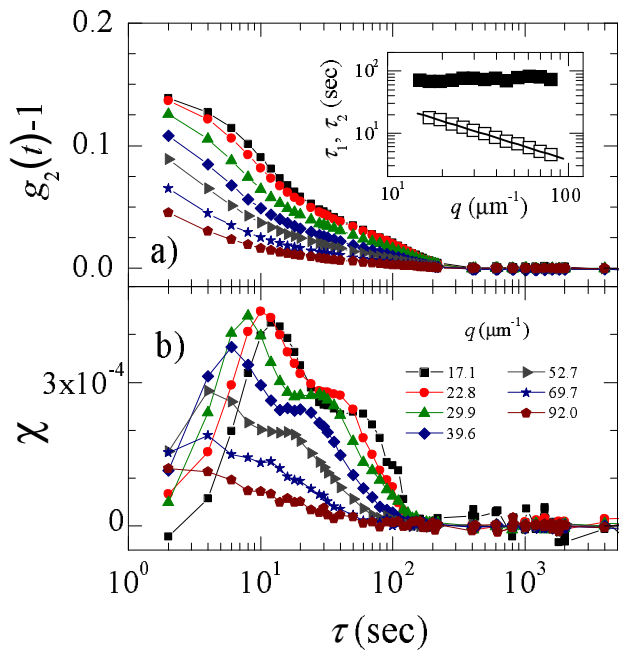}} \caption{(Color online) Top
panel, main figure: intensity correlation functions at various $q$'s
for the CB gel with $U\approx 12~k\mathrm{_{B}}T$ (data are
corrected for stray light). From top to bottom, $q$ increases from
$17.1~\mu \mathrm{m}^{-1}$ to $92.0~\mu \mathrm{m}^{-1}$ (see labels
in bottom panel). Inset: $q$-dependence of $\tau_1$ (open squares)
and $\tau_2$ (solid squares). The line is a power law fit to
$\tau_1(q)$ yielding an exponent $-0.91 \pm 0.1$. Bottom panel:
$\tau$ dependence of the dynamical susceptibility $\chi$ for the $q$
vectors corresponding to those shown in the top panel. For the sake
of clarity, not all available data have been plotted in the two main
panels.} \label{Fig:2}
\end{figure}

Figure 2a shows $g_2-1$ for the CB gel with $U\approx
12~k\mathrm{_{B}}T$, at various $q$. The intensity correlation
function exhibits a two-step decay, whose physical origin we shall
discuss later. Figure~\ref{Fig:2}b displays the $\tau$-dependent
amplitude of the noise-corrected fluctuations of the dynamics,
$\chi(\tau,q)$; at all $q$'s, $\chi$ exhibits a peaked shape,
strongly reminiscent of that observed in a variety of glassy
systems~\cite{GlotzerJChemPhys2000,DauchotPRL2005_2,KeysNaturePhysics2007,MayerPRL2004,DuriEPL2006,ChandlerPRE2006,BBBKMR,LacevicJChemPhys2003,DuriPRE2005}.
The peak of the dynamical susceptibility, $\chi^*$, occurs at a time
delay $\tau^*(q)$ of the same order of magnitude of the decay time
associated with the initial relaxation of $g_2-1$. Additionally, a
shoulder in $\chi(\tau)$ is observed on the time scale of the final
relaxation of $g_2-1$.

To gain insight on the physical origin of the slow dynamics, it
would be desirable to quantify the decay of $g_2$ by fitting its
relaxation by an appropriate function, such as the combination of
two (stretched) exponential functions. However, due to the limited
image acquisition rate, the initial decay of $g_2-1$ is only
partially captured in our experiments, especially at the largest $q$
vectors. As a consequence, a direct fit of the initial relaxation is
not reliable. As an alternative procedure, we use a scaling analysis
similar to that adopted in several photon correlation spectroscopy
works (see, e.g.,~\cite{KrallPRL1998,LucaPRL2000,BellourPRE2003}, as
shown for the gel with $U \approx 12~k\mathrm{_{B}}T$ in Fig.~3a. We
first scale the lag-time axis by $\tau^*(q)$, the lag corresponding
to the peak of the dynamical susceptibility $\chi$. This choice is
motivated by the fact that, quite generally, for glassy systems
$\tau^*$ is of the same order of magnitude of and proportional to
the system relaxation time (see
e.g.~\cite{GlotzerJChemPhys2000,LucaJPCM2003}). Furthermore, our
$\chi$ data exhibit a remarkable scaling behavior, as shown in Fig.
3b. This allows us to determine $\tau^*$ at all $q$ with an
uncertainty smaller than the temporal resolution of the CCD camera
(2 sec). We finalize our scaling procedure of the correlation
function by normalizing $g_2-1$ with a $q$-dependent amplitude,
$A(q)$, which leads to a collapse of all our $q$-dependent data in
the lag-time-range of the initial decay. To determine the shape of
the initial decay, we assume that the data can be modeled by a
stretched exponential decay, $A\exp[-(\tau/\tau_1)^{p_1}]$, and
search for the stretching exponent $p_1$ that best linearizes our
data plotted as $\ln[g_2(q,\tau)-1]-\ln A(q)$ versus
$(\tau/\tau^*)^{p_1}$, where we find an optimum value of $ p_1= 1.2
\pm 0.2$. As shown in Fig. 3a, a very good scaling onto a straight
line is obtained for $\tau \leq \tau^*$; at larger lags the data
curve upwards, because the initial decay of $g_2$ is followed by a
(tilted) plateau, as seen in Fig. 2a. The same scaling procedure is
also used to characterize the initial decay for the gel with $U
\approx 30~k\mathrm{_{B}}T$, for which we find $p_1=1.1 \pm 0.2$
(data not shown). Figure 4 shows some representative intensity
correlation functions for both gels (symbols), together with the
fits issued from the scaling analysis (lines). A full relaxation of
$g_2$ is observed only for the gel with $U \approx
12~k\mathrm{_{B}}T$, whose final decay is well approximated by a
single exponential as shown in Fig. 4b. The fact that for the gel
with $U \approx 30~k\mathrm{_{B}}T$ no final relaxation of $g_2$ is
observed in the accessible time scale is most probably due to the
deeper interparticle potential well that makes particle
displacements more difficult.

The $q$-dependence of $\tau_1$, the characteristic time of the
initial relaxation of $g_2-1$, is shown in the inset of
Fig.~\ref{Fig:2}a for the gel with $U \approx 12~k_\mathrm{B}T$. The
data can be modeled by a power law $\tau_1 \sim q^{-0.91 \pm 0.1}$,
shown in the inset of fig.~\ref{Fig:2}a as a continuous line
(similar results are also obtained for the gel with $U \approx
30~k_\mathrm{B}T$). We can exclude that the initial relaxation of
$g_2$ is due to thermally induced fluctuations of the gel branches
\cite{KrallPRL1998}, which we expect to exhibit characteristic times
at least three orders of magnitude smaller than $\tau_1$. Indeed,
both the nearly $q^{-1}$-dependence of $\tau_1$ and $p_1 > 1$
suggest that the dynamics is determined by stress-induced
rearrangements similar to those of other soft glassy materials
(see~\cite{LucaJPCM2005,BandyopadhyayReview2006} and references
therein). The final decay of the correlation function is
approximately exponential, with a decay time $\tau_2$ that is
$q$-independent. This behavior most probably stems from random rare
rearrangements occurring when bonds are broken, leading to particle
displacements larger than $1/q$. Indeed, in this case we expect one
single rearrangement to be sufficient to fully decorrelate the
contribution of the displaced scatterers to $g_2-1$. The only time
scale for the final relaxation of $g_2$ is then the average time
between rearrangements, regardless of $q$. Assuming uncorrelated
events with Poissonian statistics, one expects an exponential decay
of $g_2$, in agreement with our measurements. As mentioned before,
for the CB gel with no dispersant ($U \approx 30~k\mathrm{_{B}}T$)
no final relaxation is observed within our experimentally accessible
time window; this is most likely due to a decrease in the rate of
bond breaking as $U$ increases.

\begin{figure}
   \epsfig{file=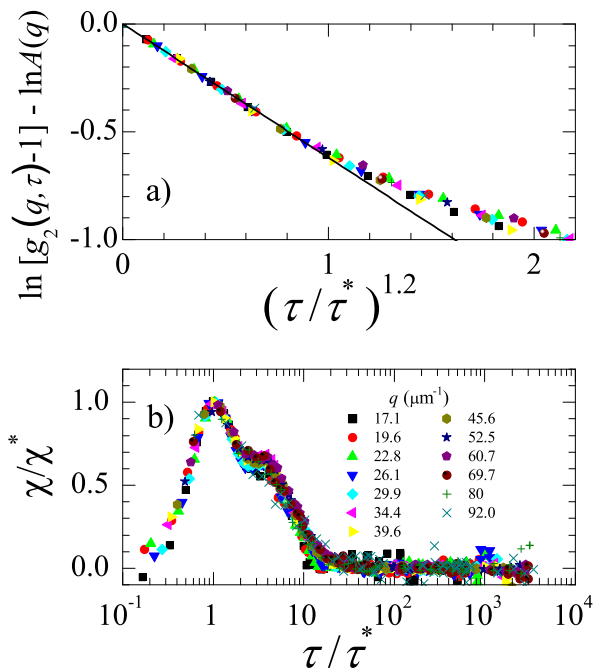,width=8cm}
 \caption{(Color online) Scaling analysis of the average dynamics and of its fluctuations, for the CB gel with $U \approx 12~k\mathrm{_{B}}T$ (same
data as in Fig. 2; here data at all the available $q$ are plotted,
except for the smallest and the largest, for which the data are more
noisy). a) Scaling plot of the initial decay
 of $g_2-1$. b) Scaling-plot of the dynamical susceptibility using reduced
 variables $\chi/\chi^*$ and $\tau/\tau^*$. $\chi^*$ and $\tau^*$
 are the height and the position of the peak of the dynamical
 susceptibility, respectively. (same symbols as in a)).}
 \label{Fig:3}.
\end{figure}

\begin{figure}
   \epsfig{file=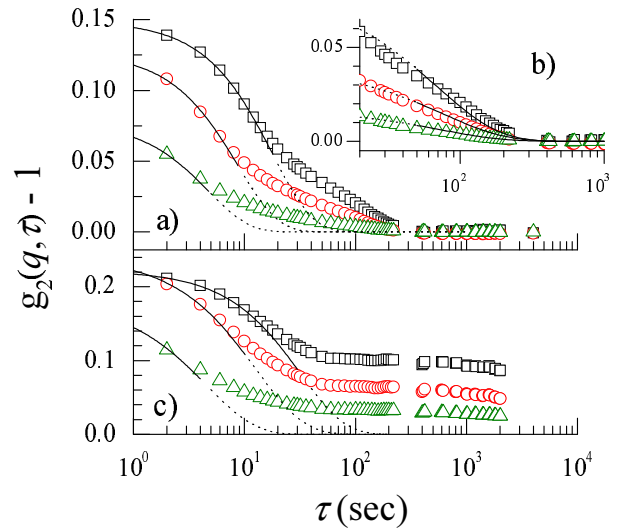,width=8cm}
 \caption{(Color online) a): Intensity correlation functions representative of the behavior at small, intermediate and large $q$
 vectors for the CB gel with 0.2\% dispersant (symbols; from top to bottom, $q = 17.1, 39.6,~\mathrm{and}~80.8~\mu\mathrm{m}^{-1}$).
 The lines are fits to the initial decay of $g_2-1$ issued from the scaling analysis described in the text.
 b): zoom on the final decay of $g_2-1$ for the same data as in a), together with the large-$\tau$ fits (lines).
 c): $g_2-1$ $vs$ $\tau$ for the CB gel with no dispersant (symbols; from top to bottom, $q = 26.5, 61.2,~\mathrm{and}~141.5~\mu\mathrm{m}^{-1}$).
 In a) and c) the fits are represented as solid lines for $\tau \leq  \tau_1$ and continued as dashed lines
 beyond the fitting interval.}
 \label{Fig:4}
\end{figure}

\begin{figure}
\centerline{\includegraphics{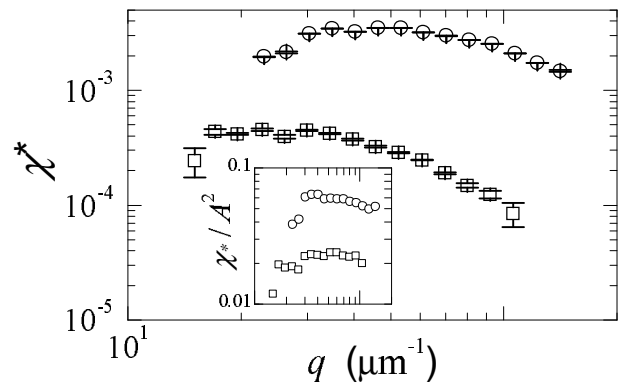}} \caption{Peak of the dynamic
susceptibility, $\chi^*$, \textit{vs} $q$ for a CB gel with 0.2\%
dispersant (squares) and no dispersant (circles). Error bars are
calculated as explained in sec.~\ref{Sec:MM}. Inset: same data
normalized by the squared amplitude of $g_2-1$.} \label{Fig:5}
\end{figure}

We investigate the length scale dependence of dynamical
heterogeneity by measuring $\chi^*(q)$, the height of the peak of
the dynamical susceptibility, shown in Fig.~5. For both gels, we
find that $\chi^*(q)$ first increases with $q$, reaches a maximum
value at $q \equiv q^*$ and then decreases at larger scattering
vectors. Moreover, we find that the dynamical fluctuations are more
pronounced for the gel with the deeper interparticle potential well.
This can be intuitively understood as the size of the regions that
rearrange cooperatively is presumably larger in gels with stronger
interparticle interactions, leading to larger dynamical
fluctuations. Additionally, it is conceivable that the relaxation
process itself is more heterogeneous in time for stronger gels,
where internal stress build-up and release is likely to be more
important, further contributing to enhanced fluctuations. As
outlined in ref.~\cite{DuriEPL2006}, $\chi$ should in principle be
normalized by the ($q$-dependent) squared amplitude of the
relaxation of $g_2-1$, to allow for an explicit comparison of data
obtained at different $q$'s.~\cite{noteqdependence}. Here, such a
correction is affected by a large uncertainty, since the short-time
behavior of $g_2$ is barely accessible to the CCD. The height of the
peak of the variance normalized using the amplitude estimated via
the scaling procedure, $\chi^*/A^2$, is shown in the inset of
Fig.~\ref{Fig:5}. Although the data are more noisy, the peaked shape
observed in the main plot is preserved.

\section{Discussion}
\label{Sec:D}

\begin{figure}
\centerline{\includegraphics{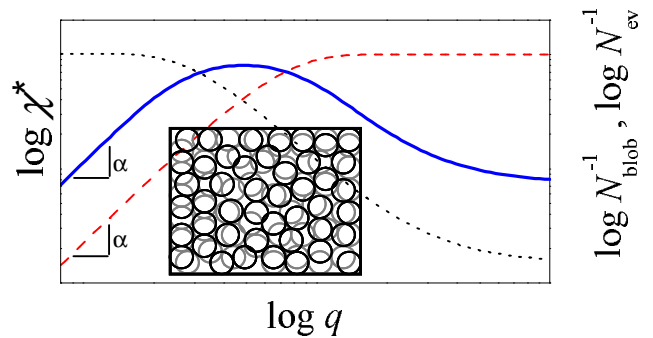}}  \caption{(Color online)
Inset: schematic representation of the rearrangement events within
the scattering volume. Successive events at the same location are
indicated by circles of different grey level. Main figure:
qualitative double logarithmic plot of the proposed $q$-dependence
of dynamical fluctuations. Left axis: $\chi^*$ (continuous line,
blue); right axis: $\Ne^{-1}$ (dashed line, red), and $\Nb^{-1} \sim
\Vb$ (dotted line, black). The slope $\alpha$ varies between 1 and
2, depending on the nature of the dynamics, as discussed in the
text.} \label{Fig:6}
\end{figure}

The $q$ dependence of $\chi^*$ found here is very different from
that measured in low-$q$ light scattering experiments on strongly
attractive gels made of spherical polystyrene
particles~\cite{DuriEPL2006}, where $\chi^* \sim q$. One may wonder
whether this discrepancy stems from a difference in the particle
morphology or polydispersity. However, CB suspensions have been
found to exhibit the same relaxation spectrum (as measured by
rheology) as model systems consisting of spherical, nearly
monodisperse colloids interacting via short-ranged depletion
forces~\cite{PrasadFaraday2003}. Thus, the dynamical properties
reported here are likely to be generic for colloidal gels with
moderate attractive interactions. Instead, we propose that the
different $q$-dependence of $\chi^*$ arises from the different $q$
and $U$ ranges probed in this work compared to~\cite{DuriEPL2006}.
In this section, we introduce a simple
---yet general--- scaling argument for the length-scale dependence
of dynamical fluctuations in glassy systems that reconciles these
contrasting observations. We assume the dynamics to be due to random
rearrangement events each affecting a ``blob'' of
volume $\Vb$. 
The fluctuations of $c_I$ then stem from fluctuations of $N_{\mathrm
{tot}}$, the total number of events needed to decorrelate the
scattered light. In determining $N_{\mathrm {tot}}$, one has to take
into account that one single event may not displace particles far
enough to fully suppress the local contribution to the scattered
light; additionally, one single event may affect only a portion of
the whole scattering volume. Thus, $N_{\mathrm {tot}} \sim \Nb \Ne$,
where $\Nb \sim V_{\mathrm{sc}}/\Vb$ is the number of dynamically
correlated blobs of volume $\Vb$ contained in the scattering volume
$V_{\mathrm{sc}}$, and $\Ne$ is the number of rearrangement events
that are needed, at any given location, to relax the local
contribution to the correlation function, i.e. the number of events
on the time scale $\tau_\mathrm{r}$ of the system's relaxation. The
inset of Fig.~\ref{Fig:6} is a schematic representation of this
concept for $\Nb \sim 50$ and $\Ne=2$: in this case, $g_2-1$ decays
to zero when at least two rearrangement events (symbolized by a
black and a gray circle) have occurred at every location in the
system. Given the random nature of the events, $N_{\mathrm {tot}}$
fluctuates, and so does $c_I$, with $\mathrm{var}(c_I) \sim
\mathrm{var}(N_{\mathrm {tot}})$. According to the central limit
theorem, the relative variance of $N_{\mathrm {tot}}$ ---and thus
$\chi$--- is expected to scale as $N_{\mathrm {tot}}^{-1} \sim (\Nb
\Ne)^{-1}$.

We stress that in general both $\Nb$ and $\Ne$ are $q$-dependent
quantities. Indeed, $\Ne$ decreases as $q$ increases, because fewer
events are needed to displace the particles over the smaller
distances corresponding to larger scattering vectors. More
specifically, in our model the average number of events is
proportional to time and thus we expect $\Ne \sim \tau_{\mathrm{r}}
\sim q^{-\alpha}$. For uncorrelated particle displacements due to
successive events (``Brownian-like'' rearrangements), $\alpha = 2$.
By contrast, $\alpha = 1$  if the displacement direction persists
over several events (``ballistic-like'' rearrangements), as found
for our CB gels and other systems with internal stress-driven
dynamics \cite{LucaJPCM2005,DuriEPL2006,BandyopadhyayReview2006}.
Note that at very large $q$ $\Ne$ should saturate to one, when the
particles' displacement due to one single event exceeds $1/q$, the
length scale probed in a scattering experiment. By contrast, there
are no \textit{a priori} prescriptions on the $q$-dependence of
$\Nb$. However, it is reasonable to assume that the latter be an
increasing function of $q$, since in glassy systems rearrangements
involving displacements over large distances (probed at low $q$) are
likely to require the highly cooperative motion of many particles,
whereas smaller displacements (probed at high $q$) may be achieved
independently by clusters containing just a few particles. Thus, for
$q \rightarrow 0$, $\Nb$ is expected to saturate at a lower bound,
which may be as low as 1 when $\Vb$ is larger than the scattering
volume, as observed in~\cite{DuriEPL2006}. In the opposite limit, $q
\rightarrow \infty$, $\Nb$ should saturate to $N_{\mathrm p}$, the
number of particles in the system.

The main panel of Fig.~\ref{Fig:6} illustrates schematically the
$q$-dependence of $\Nb^{-1}$ and $\Ne^{-1}$ (right axis) and that of
$\chi^* \sim (\Nb \Ne)^{-1}$ (left axis). At low $q$, $\chi^* \sim
q^{\alpha}$, since $\Nb^{-1}$ saturates to one whereas $\Ne^{-1}
\sim q^{\alpha}$. This is the regime observed for the strongly
attractive gels of ref.~\cite{DuriEPL2006}, for which $\Nb = 1$ and
$\chi^* \sim \Ne^{-1} \sim \tau_r(q)^{-1} \sim q$ ($\alpha=1$). By
contrast, $\chi^*$ tends to a constant value at very large $q$,
because $\Nb^{-1}$ saturates to $N_{\mathrm p}^{-1}$, while
$\Ne^{-1}$ saturates to one. The behavior of the dynamical
susceptibility for intermediate $q$'s depends on the detailed
interplay between $\Nb$ and $\Ne$. In Fig.~\ref{Fig:6} we have
sketched the case where $\Nb^{-1}$ decreases faster than $\Ne^{-1}$
grows, yielding a peaked shape of $\chi^*$, as observed in the
experiments presented here.

We expect these scaling arguments to hold also for other glassy
systems exhibiting dynamical fluctuations. Indeed, a growing trend
in the low $q$ regime, with $\alpha=2$, is predicted for dynamically
facilitated models, for which $\chi^*$ saturates in the opposite
limit $q\rightarrow \infty$~\cite{ChandlerPRE2006}. For a 2-D
granular system~\cite{DauchotPRL2005_2} and in simulations of a
Lennard-Jones glass former~\cite{LacevicJChemPhys2003} and a
short-range attractive glass~\cite{CharbonneauCondmat2007}, a
non-monotonic behavior similar to that of the CB gels has been
reported, with $\chi^*$ initially growing with $q$ but then
decreasing after going through a peak. The length scale
corresponding to this peak is comparable to the inter-particle
distance for repulsive systems, while it is shifted to smaller
values, corresponding to the width of the interparticle potential
well, for attractive glasses~\cite{CharbonneauCondmat2007}. Finally,
the high-$q$ regime has been explored for hard spheres within the
mode coupling theory in~\cite{BBBKMR}, where the amplitude of
$\chi_{\phi}$, a lower bound for $\chi_4$, was reported to decrease
with $q$ around the peak of the static structure factor, in
agreement with the behavior predicted by our scaling argument at
large $q$.

Simulations of glass formers and experiments on 2D granular
materials suggest that $\chi^*$ is maximum on the length scale of
the particle size or that of the interparticle bond. Colloidal gels
present an additional characteristic length, the size of the
(fractal) clusters that compose them. For the gels of
ref.~\cite{DuriEPL2006}, $\chi^*$ was shown to grow with $q$ for
length scales intermediate between the cluster and the particle
size. For the CB gels studied here, $\chi^*$ peaks at $q^*\approx 20
- 50~\mu \mathrm{m}^{-1}$ (see Fig.~\ref{Fig:5}), corresponding to a
length scale $\Lambda \equiv 2\pi/q^* \approx 125 -
300~\mathrm{nm}$, comparable to the particle size. Collectively,
these observations suggest that the crossover length scale for
dynamical heterogeneity in colloidal gels is of the order of or
smaller than the particle size, similarly to molecular glass formers
and granular materials, rather than the cluster size. Interestingly,
in our gels $\Lambda$ shifts towards smaller values ($q^*$
increases) when the particles are more sticky, the same trend as
that reported in Ref.~\cite{CharbonneauCondmat2007} when going from
a nearly hard-sphere system to an attractive one. Although the exact
value of $q^*$ most likely depends on the detailed shape of the
interparticle potential, it is intriguing to note that the values
found here ($q^*R \approx 3.6$ for $U \approx 12 k\mathrm{_{B}}T$
and $q^*R \approx 9$ for $U \approx 30 k\mathrm{_{B}}T$,
respectively) are comparable to those shown
in~\cite{CharbonneauCondmat2007} for a nearly-hard-sphere and an
attractive system ($q^*R \approx 2.5$ and $q^*R \approx 6$,
respectively).

In conclusion, we have shown that the dynamics of CB gels is
temporally heterogeneous. Dynamical fluctuations increase with $q$,
peak around the inverse particle size, and decrease at larger
scattering vectors. This behavior and that of other systems can be
rationalized by a simple scaling argument, providing a general
framework for understanding temporal fluctuations of the dynamics in
glassy systems. Additionally, our measurements demonstrate that XPCS
may be used to obtain quantitative information not only on the
average dynamics, but also on its heterogenous behavior. This opens
a new way to investigate dynamical heterogeneity in a wide variety
of materials, possibly including molecular glass formers, whose
characteristic length scales match those probed by XPCS, provided
that a high enough signal can be collected.

We thank ESRF for provision of synchrotron radiation facilities and
financial support. We are indebted to P. Clegg for generously
providing us with some of his beam time at ESRF for control
experiments. We thank L. Berthier, W. Kob and A. Duri for many
useful discussions. This work was supported in part by the European
MCRTN ``Arrested matter'' (MRTN-CT-2003-504712), the NoE
``SoftComp'' (NMP3-CT-2004-502235), the Swiss National Science
Foundation, the French CNRS (PICS no. 2410), and ACI JC2076 and ANR
JCJC-CHEF grants. L.C. is a junior member of the Institut
Universitaire de France, whose support is gratefully acknowledged.



\begin{references}
\bibitem{notenormalization} Note that, contrary to $\chi_4$ in
simulations, $\chi^{\mathrm{(exp)}}(\tau,q)$ is not normalized with
respect to $N_{\mathrm{p}}$, the number of probed particles. For the
two gels studied here, $N_{\mathrm{p}}$ is the same, because
$\varphi$ does not change.
\end{references}
\end{document}